\documentclass[a4paper,11pt]{article}
\usepackage{pos}
\usepackage{subcaption}
\usepackage{bm,bbm}

\let\oldbibliography\thebibliography
\renewcommand{\thebibliography}[1]{%
  \oldbibliography{#1}%
  \setlength{\itemsep}{1.8pt}%
  \setlength{\parskip}{0pt}%
}

\makeatletter
\g@addto@macro\bfseries{\boldmath}
\makeatother

\title{Three-body study of the $T_{cc}(3875)^+$ from lattice QCD}

\author[a]{Herzallah~Alharazin}
\author*[a]{Andr\'e~Bai\~{a}o~Raposo}
\author[a]{John~Bulava}
\author[b1,b2]{Sebastian~Dawid}
\author[c]{Jeremy~R.~Green}
\author[d]{Colin Morningstar}
\author[e]{Fernando~Romero-L\'opez}
\author[e]{Miguel~Salg}
\author[f]{Stephen~R.~Sharpe}
\author[g]{Andres~Stump}

\affiliation[a]{Institut f\"ur Theoretische Physik II, Ruhr-Universit\"at Bochum,\\
  D-44780 Bochum, Germany}
\affiliation[b1]{Physics  Department,  Indiana  University, Bloomington,  IN  47405,  USA}
\affiliation[b2]{Center for  Exploration  of  Energy  and  Matter,  Indiana  University,  Bloomington, IN, 47403, USA}
\affiliation[c]{John von Neumann-Institut für Computing NIC,
  Deutsches Elektronen-Synchrotron DESY,\\
Platanenallee 6, 15738 Zeuthen, Germany}
\affiliation[d]{Department of Physics, Carnegie Mellon University, \\
    Pittsburgh, Pennsylvania 15213, USA}
\affiliation[e]{Institute for Theoretical Physics, 
	Albert Einstein Center for Fundamental Physics, \\ 
	University of Bern, 3012 Bern, Switzerland}
\affiliation[f]{Physics Department, University of Washington, \\
Seattle, WA 98195-1560, USA}
\affiliation[g]{Institut für Physik, Humboldt-Universität zu Berlin,\\
  Zum Großen Windkanal 2, 12489 Berlin, Germany}

\emailAdd{herzallah.alharazin@ruhr-uni-bochum.de}
\emailAdd{andre.baiaoraposo@ruhr-uni-bochum.de}
\emailAdd{john.bulava@ruhr-uni-bochum.de}
\emailAdd{sdawid@iu.edu}
\emailAdd{jeremy.green@desy.de}
\emailAdd{fernando.romero-lopez@unibe.ch}
\emailAdd{miguel.salg@unibe.ch}
\emailAdd{srsharpe@uw.edu}
\emailAdd{andres.stump@hu-berlin.de}

\abstract{We discuss an ongoing first lattice study of the doubly-charmed tetraquark $T_{cc}^+$(3875) via a three-body approach. We investigate the $DD\pi$ system in the $I=0$, $C=2$ sector, where the $T_{cc}^+$ appears as a pole in the $J^P = 1^+$ $DD\pi$ elastic scattering amplitude. The approach automatically incorporates two-body $DD^*$ and three-body $DD\pi$ effects and treats left-hand cuts due to single $\pi$ exchanges. Two CLS ensembles, X252 and X253, with pion mass $M_\pi \approx 280$~MeV, are used, and an operator set comprised of two- and three-hadron and tetraquark operators is employed to extract finite-volume energies. Additional inputs are required for the three-body finite-volume analysis, in the form of amplitudes for the $I=1$ $DD$ and $I=1/2$ $D\pi$ two-body subsystems. We present preliminary results for these subchannels and perform exploratory three-body spectra determinations for simple choices of the three-particle K-matrix $\mathcal{K}_{\rm df, 3}$, allowing a first comparison to the lattice spectrum.}

\FullConference{The 42nd International Symposium on Lattice Field Theory (LATTICE2025)\\
2-8 November 2025\\
Tata Institute of Fundamental Research, Mumbai, India\\}

\renewcommand{\hookAfterAbstract}{%
  \par\vspace{5pt}
  \hfill DESY-26-021, HU-EP-26/10-RTG
  \par\vspace{20pt}
}

\begin{document}

\maketitle

\section{Introduction}
\label{sec:intro}

The multitude of exotic hadrons discovered in recent decades has led to significant efforts to understand their properties directly from the underlying theory of the strong interactions, quantum chromodynamics (QCD). A noteworthy example is the doubly-charmed tetraquark, $T_{cc}(3875)^+$, observed by the LHCb collaboration in 2021 \cite{LHCb:2021vvq, LHCb:2021auc}. This state, with quantum numbers $I(J^P) = 0(1^+)$ and minimal quark content $cc\bar u \bar d$, is a very narrow resonance decaying primarily to three-hadron states via intermediate two-hadron states, $T_{cc}^+ \to D D^* \to DD\pi$. Its structure is still a matter of debate -- the proximity to the $DD^*$ threshold suggests a meson-meson molecule, but compact tetraquark or diquark-antidiquark structures have also been suggested \cite{Hanhart:2025bun}.

Lattice QCD can help clarify the properties of the $T_{cc}^+$ from first principles, and indeed several calculations have already investigated this state \cite{Padmanath:2022cvl, Chen:2022vpo, Lyu:2023xro, Collins:2024sfi, Whyte:2024ihh, Prelovsek:2025vbr}. These works have used heavier-than-physical pion masses, for which the $D^*$ meson is stable and the $T_{cc}^+$ can be studied through the simpler two-body process of $DD^*$ scattering. The standard lattice approach to two-body scattering relies on the L\"uscher formalism, which maps finite-volume lattice energies to infinite-volume scattering observables. However, in systems like $DD^*$, this method is limited because of the so-called left-hand cuts \cite{Du:2023hlu}, subthreshold branch cuts in the partial-wave-projected amplitude, arising here due to $u$-channel single-pion exchanges between $D$ and $D^*$. Such cuts invalidate the formalism when applied to lattice energies near to or on the cut, complicating the determination of the $DD^*$ amplitude in the subthreshold region, where $T_{cc}^+$ poles are expected to appear.

Various proposals have been made to address the left-hand cut issue \cite{Raposo:2023oru, Meng:2023bmz, Hansen:2024ffk, Bubna:2024izx, Dawid:2024oey, Raposo:2025dkb}
and analyses of $DD^*$ data taking into account the effect of the pion exchange already exist \cite{Meng:2023bmz, Collins:2024sfi, Prelovsek:2025vbr, Dawid:2024dgy}. In this talk, we employ one such proposal, i.e.~the relativistic field-theoretic (RFT) three-body formalism on the isospin-0 $DD\pi$ system, as laid out in ref.~\cite{Hansen:2024ffk} and first explored in ref.~\cite{Dawid:2024dgy}. This method not only avoids the left-hand cut problem entirely, but naturally accommodates both $DD^*$ and three-particle $DD\pi$ states. Since the $T_{cc}^+$ decays to a $DD\pi$ final state in nature, three-body states become increasingly relevant as we move towards physical pion masses. Therefore, this approach offers a flexible method to investigate the $T_{cc}^+$ from lattice QCD across a range of pion masses.

We start by giving a short overview of the three-body RFT formalism and how it can be used to study the $T_{cc}^+$ (section \ref{sec:formalism}). We then describe the lattice setup and the extraction of finite-volume energies for the $DD^* + DD\pi$ system and the $DD$ and $D\pi$ subsystems (section~\ref{sec:setup}), and present the preliminary L\"uscher analyses of the two-body subchannels and initial fits to the $DD^* + DD\pi$ spectrum using the three-body quantization condition (section~\ref{sec:results}).


\section{Overview of the RFT approach}
\label{sec:formalism}

We briefly outline the RFT three-body formalism for the $T_{cc}^+$ below, following ref.~\cite{Hansen:2024ffk}. 
In this approach, the $DD\pi$ system is split into a spectator and a pair, leading to two distinct two-particle subsystems depending on the choice of spectator flavor: $I=1/2$ $D\pi$ for a $D$ spectator, and $I=1$ $DD$ for a $\pi$ spectator. The key idea is then that the vector $D^*$ meson can be encoded as a pole in the $p$-wave amplitude of the $D\pi$ subsystem, which allows the formalism to simultaneously handle $DD\pi$ and $DD^*$ states. The pion mass determines whether the $D^*$ is a bound state or resonance. An additional state to consider is the scalar  $D_0^*(2300)$, which contributes to the $s$-wave $D\pi$ amplitude.

\subsection{Finite-volume formalism}
\label{sec:FV}

The centerpiece of the formalism is the three-particle quantization condition (QC3), which is satisfied at the finite-volume energies of the $DD^* + DD\pi$ system. For a cubic box of side $L$ with periodic boundary conditions, it takes the form
\begin{equation}
\det_{i\boldsymbol k \ell m} \left[ \widehat{\mathcal K}_{\rm df,3}(E^\star) + \widehat{F}_3(E,\boldsymbol P, L)^{-1} \right] = 0 \,,
\label{eq:QC3_def}
\end{equation}
where $(E,\boldsymbol P)$ are the total energy and three-momentum and $E^\star = \sqrt{E^2 - \boldsymbol P{}^2}$ is the center-of-mass energy. 
The objects in the determinant are matrices in a space with indices $i\boldsymbol k \ell m$:~the first two indices, $i \in \{D, \pi\}$ and $\boldsymbol k \in \frac{2\pi}{L} \mathbb Z^3$, specify the flavor and momentum of the spectator particle, while $\ell$ and $m$ specify the orbital angular momentum of the remaining $DD$ or $D\pi$ pair in its rest frame. $\widehat{\mathcal K}_{3, \rm df}$ is directly related to the three-particle K-matrix $\mathcal K_{3,\rm df}$, a scheme-dependent infinite-volume object parameterizing the short-range three-body physics. The matrix $\widehat{F}_3$ is explicitly given by
\begin{align}
\widehat{F}_3 \equiv \frac{\widehat{F}}{3} - \widehat{F} \frac{1}{1 + \widehat{\mathcal M}_{2,L} \widehat{G}} \widehat{\mathcal M}_{2,L} \widehat{F} \,, \qquad \widehat{\mathcal M}_{2,L} \equiv \frac{1}{\widehat{\mathcal K}_{2,L}^{-1} + \widehat{F}} \,,
\label{eq:F3_def}
\end{align}
where $\widehat{F}$, $\widehat{G}$ are kinematic functions ($\widehat{F}$ is closely related to the multichannel L\"uscher zeta function and $\widehat{G}$ effects swaps of the spectator particle) and $\widehat{\mathcal K}_{2,L}$ packages information about the dynamics of the two-particle subsystems, through their K-matrices. The definitions of these objects are given in full in Section 3.2 of ref.~\cite{Dawid:2024dgy}.
Parameters for the two-particle K-matrices can be fixed from separate lattice or experimental values, as done in ref.~\cite{Dawid:2024dgy}. In the current study, we instead extract these by direct fits to the two-body finite-volume spectra, obtained on the same ensembles as the $DD^* + DD\pi$ system.

Combining the subchannel information with the QC3, we can then constrain $\mathcal K_{\rm df, 3}$ from the $DD^* + DD\pi$ spectrum. We use a threshold expansion to parametrize $\mathcal K_{\rm df, 3}$~\cite{Blanton:2021mih}:
\begin{equation}
\mathcal K_{\rm df, 3} (\{\boldsymbol k'\}, \{\boldsymbol k\}) = \mathcal K_3^{\rm iso,1} + \mathcal K_3^{\rm iso,2} \Delta +  \mathcal K_3^B \Delta_2^S + \mathcal K_3^E \Tilde{t}_{22} \,,
\label{eq:K3_thr}
\end{equation}
where $\{\boldsymbol k\}, \{\boldsymbol k'\}$ are the sets of initial and final state three-momenta, respectively, and $\mathcal K_3^{\rm iso,1}$, $\mathcal K_3^{\rm iso,2}$, $\mathcal K_3^B$ and $\mathcal K_3^E$ are real parameters. The kinematic variables $\Delta$, $\Delta_2^S$ and $\Tilde{t}_{22}$ are defined as
\begin{align}
\Delta \equiv \frac{E^2 - M^2}{M^2} \,, \qquad
\Delta_2^S \equiv \frac{\sigma_{DD}' + \sigma_{DD} - 8M_D^2}{M^2} \,, \qquad
\Tilde{t}_{22} \equiv \frac{(k'_\pi - k_\pi)^2}{M^2} \,,
\label{eq:K3_kine}
\end{align}
where $M \equiv 2M_D + M_\pi$ is the $DD\pi$ threshold energy, $\sigma_{DD}, \sigma_{DD}'$ are the two-body invariant masses squared for the $DD$ subsystem in the initial and final state, while $k_\pi, k_\pi'$ are the initial and final four-momenta of the pion.

\subsection{From \ensuremath{\mathcal K_{\rm df, 3}} to the \ensuremath{DD^*} amplitude}
\label{sec:amplitudes}

Given the two- and three-body K-matrices, we determine the connected $DD\pi$ amplitude, $\boldsymbol{\mathcal M}_3$, by numerically solving the RFT integral equations. 
We use a compact matrix notation below, with multiplication signifying integration over intermediate loop momenta. All objects are projected to definite $J^P$ and expressed in the total orbital angular momentum and spin $L S$ basis, so that the three-body amplitude carries channel and partial-wave indices. $\boldsymbol{\mathcal M}_3$ is given by
\begin{equation}
\bm{\mathcal{M}}_3
=
\bm{\mathcal{D}}
+
\bm{\mathcal{M}}_{\mathrm{df},3}\,,
\end{equation}
where $\bm{\mathcal{D}}$ resums one-particle exchanges, and $\bm{\mathcal{M}}_{\mathrm{df},3}$ includes all short-range three-body interactions. The ladder amplitude satisfies an integral equation, which can be written schematically as
\begin{equation}
\bm{\mathcal{D}}
=
-
\bm{\mathcal{M}}_2\,\bm{\mathcal{G}}\,\bm{\mathcal{M}}_2
-
\bm{\mathcal{M}}_2\,\bm{\mathcal{G}}\,\bm{\mathcal{D}}\,,
\end{equation}
where
$\bm{\mathcal{G}}$ is the single-pion exchange and $\bm{\mathcal{M}}_2$ includes the $DD$ and $D\pi$ amplitudes, specified through their K-matrices.
The three-body K-matrix $\mathcal K_{\rm df, 3}$ enters through the \textit{divergence-free amplitude}, $\bm{\mathcal{M}}_{\mathrm{df},3}$, which obeys
\begin{equation}
\bm{\mathcal{M}}_{\mathrm{df},3}
=
\bm{\mathcal{L}}\,
\bm{\mathcal{T}}\,
\bm{\mathcal{R}}\,, 
\qquad
\bm{\mathcal{T}}
=
\bm{\mathcal{K}}_{\rm df, 3}
-
\bm{\mathcal{K}}_{\rm df, 3}\,\bm{\tilde{\rho}}\,
\bm{\mathcal{L}}\,
\bm{\mathcal{T}}\,.
\end{equation}
where $\bm{\mathcal{L}}$ and $\bm{\mathcal{R}}$ resum external two-body rescatterings and $\bm{\tilde{\rho}}$ is a modified phase-space factor. 
Detailed definitions are given in section 2 of ref.~\cite{Dawid:2024dgy}.

The $DD^*$ amplitude is obtained from the $DD\pi$ solution via LSZ reduction at the $D^*$ pole of the $D\pi$ $p$-wave amplitude. Near the pole, the $p$-wave $D\pi$ amplitude is, approximately,
\begin{equation}
\mathcal{M}_{2,\,p\text{-wave}}^{(D\pi)}
\;\sim\;
\frac{\zeta^2}{\sigma_{D\pi} - M_{D^*}^2}\,,
\label{eq:Dpi_amp_pole}
\end{equation}
where $\sigma_{D\pi}$ is the two-body invariant mass squared. Amputating the external pole factors of the $DD\pi$ amplitude and setting the external $D\pi$ invariant masses to the $D^*$ mass yields the required $DD^*$ amplitude. 


\section{Lattice setup and spectrum determination}
\label{sec:setup}

This work was based on two CLS ensembles, X252 and X253; see Table \ref{tab:ensembles}. Both are $N_f = 2+1$ ensembles with $O(a)$-improved Wilson fermions, with the same lattice spacing and pion mass $M_\pi \approx 280$ MeV, but different volumes. The valence charm quark mass was tuned such that the $D$ mass is the average of the physical $D^0,D^+,D_s^+$ masses. The analysis presented here has primarily used the X252 ensemble, with X253 used at this stage only for the $I = 1/2$ $D\pi$ amplitude.

\begin{table}[h]
\centering
\begin{tabular}{c c c c c c c}
\hline\hline
 {} & $L/a$ & $T/a$ & $M_\pi L$ & $M_\pi[\rm MeV]$ & $a[\rm fm]$ & $N_{\rm LapH}$ \\
\hline\hline
X252 & 36 & 128 & 3.4 & 280 & 0.064 & 48 \\
X253 & 40 & 128 & 3.7 & 280 & 0.064 & 64 \\
\hline
\end{tabular}
\caption{CLS ensembles used in this work.}
\label{tab:ensembles}
\end{table}

The operator bases were built from local single-hadron operators projected to definite momentum and products of two or three such operators, combined to transform under the finite-volume irreps of interest. Frames with total momentum $(L\boldsymbol P/2\pi)^2 = 0,1,2,3$ and individual particle momenta going up to $(L\boldsymbol p/2\pi)^2 = 3$ were considered. For $I = 1/2$ $D\pi$, the operator set includes $D^*$ and  $D_0^*$ operators and bilocal $D\pi$ and $D^*\pi$. For $I = 1$ $DD$, we have bilocal $DD$, $D^*D$, $D^*D^*$ and $D_0^*D_0^*$ operators. Finally, for the main $DD\pi + DD^*$ system, bilocal $DD^*$, $D^*D^*$, $D_0^* D$, $D_0^* D^*$ and trilocal $DD\pi$ operators are used, in addition to local $DD^*$ and $D^*D^*$ tetraquark operators, e.g.
\begin{equation}
T^{DD^*}_i (t, \boldsymbol p)  = \sum_{\boldsymbol x} e^{-i\boldsymbol p \cdot \boldsymbol x} \left[ \bar u\gamma_5 c \, \bar d \gamma_i c - \bar d \gamma_5 c \, \bar u \gamma_i c \right] (t,\boldsymbol x) \,.
\label{eq:Tcc_ops}
\end{equation}

The distillation framework~\cite{HadronSpectrum:2009krc} is used to compute correlator matrices. Perambulators were computed using the QUDA-LapH package \cite{quda_laph} (number of LapH eigenvectors shown in Table \ref{tab:ensembles}). To compute correlators involving tetraquark operators, the position-space sampling of ref.~\cite{Stump:2025owq} was employed. A GEVP is performed on the correlator matrix to extract the finite-volume energy levels. 

For single hadrons and the $D^*$- and $D_0^*$-like levels in the $D\pi$ system, the energy extraction used single-exponential fits directly to the correlators. For all other levels, fits were performed to ratios of correlators, where the two- or three-hadron correlators are divided by the product of single-hadron correlators with momentum assignments chosen based on the overlaps to the operators, to extract energy shifts. Interacting energies are then reconstructed using continuum dispersion relations and the extracted single-hadron masses.

We emphasize the importance of tetraquark operators, which significantly impact several energy levels, as pointed out in recent works \cite{Prelovsek:2025vbr, Stump:2025owq}, and note also the relevance of $DD\pi$ operators to resolve lower-lying $DD\pi$-like states, e.g.~the ground state in the rest frame $A_{1u}$ irrep.


\section{Preliminary results}
\label{sec:results}

This preliminary analysis consists of initial checks on each of the three relevant systems. The $DD$ and $D\pi$ subsystems were investigated individually, with a finite-volume analysis performed on each to select reasonable K-matrix parameterizations and determine best fit parameters. The main $DD^* + DD\pi$ system was then studied, using the QC3 to fit the spectrum and determine the three-particle K-matrix, fixing the two-body parameters to those obtained from the subsystems.

\subsection{Isospin-1 \ensuremath{DD} subsystem}
\label{sec:DD_fitting}

\begin{figure}[h]
\centering
\includegraphics[width=0.6\linewidth]{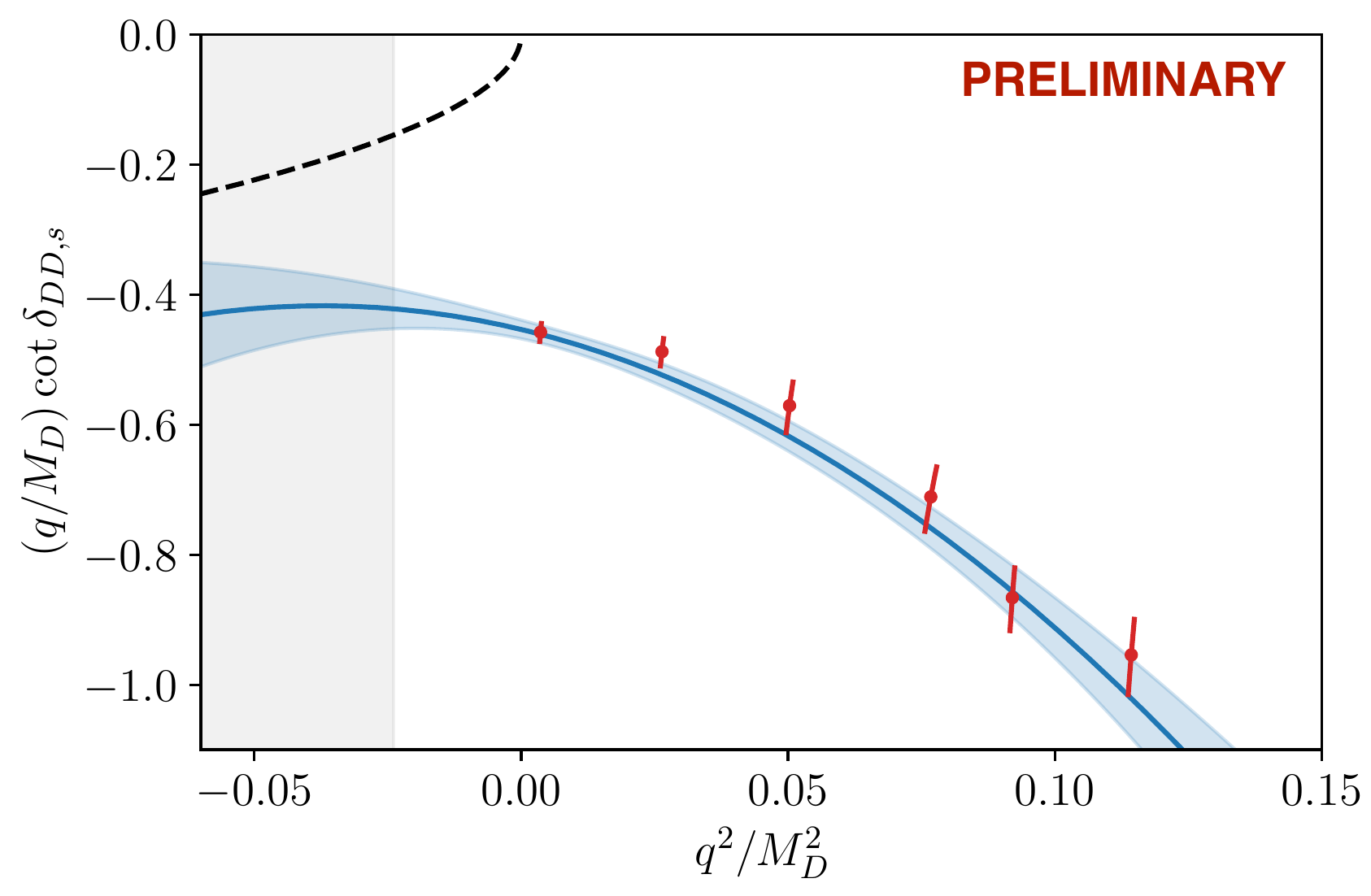}
\caption{Plot of the $s$-wave phase shift for the $I=1$ $DD$ subsystem. The solid blue line is the best ERE3 fit described in the text and the red points show the data. Two levels included in the fit (ground states of the $(L\boldsymbol P/2\pi)^2=2,3$ frames) are not shown, as their error bars cross free-energy singularities of the zeta function. The dashed black line shows the bound state condition and the shaded gray region corresponds to the left-hand cut due to two-pion exchange.}
\label{fig:DD_fit}
\end{figure}

To constrain the $s$-wave amplitude for the $I=1$ $DD$ system, eight levels were obtained on the X252 ensemble across the $A_{1g}(0)$, $A_1(1)$, $A_1(2)$ and $A_1(3)$ irreps (the number in brackets corresponds to the value of $(L\boldsymbol P/2\pi)^2$). All eight levels lie below the $D^*D^*$ threshold and were included in the fit. Two- and three-term effective range expansions (ERE2 and ERE3 for short) were used to parameterize the $s$-wave phase shift. An ERE3 parametrization of the form
\begin{equation}
\frac{q}{M_D} \, \cot \delta_{DD,s} = C_1 + C_2 \left(\frac{q}{M_D}\right)^2 + C_3 \left(\frac{q}{M_D}\right)^4 \,,
\label{eq:DD_ERE3}
\end{equation}
where $M_D$ is the $D$ meson mass, $q$ is the center-of-mass scattering momentum and $\delta_{DD,s}$ is the phase shift, was chosen. The best fit parameters obtained were
\begin{equation}
C_1 = -0.454(14) \,, \quad C_2 = -1.96(77) \,, \quad C_3 = -26.3(8.5) \,,
\label{eq:DD_ERE3_params}
\end{equation}
leading to $\chi^2/{\rm d.o.f.} = 5.58/5 = 1.17$. These yield a scattering length and effective range of $a_{0,s} M_D = 2.21(7)$ and $r_{0,s} M_D = -3.91(1.54)$, confirming the expectation of a relatively weak interaction. The fit results are shown in Figure \ref{fig:DD_fit}.

\subsection{Isospin-1/2 \ensuremath{D\pi} subsystem}
\label{sec:Dpi_fitting}

\begin{figure}[h]
\centering
\includegraphics[width=1.\linewidth]{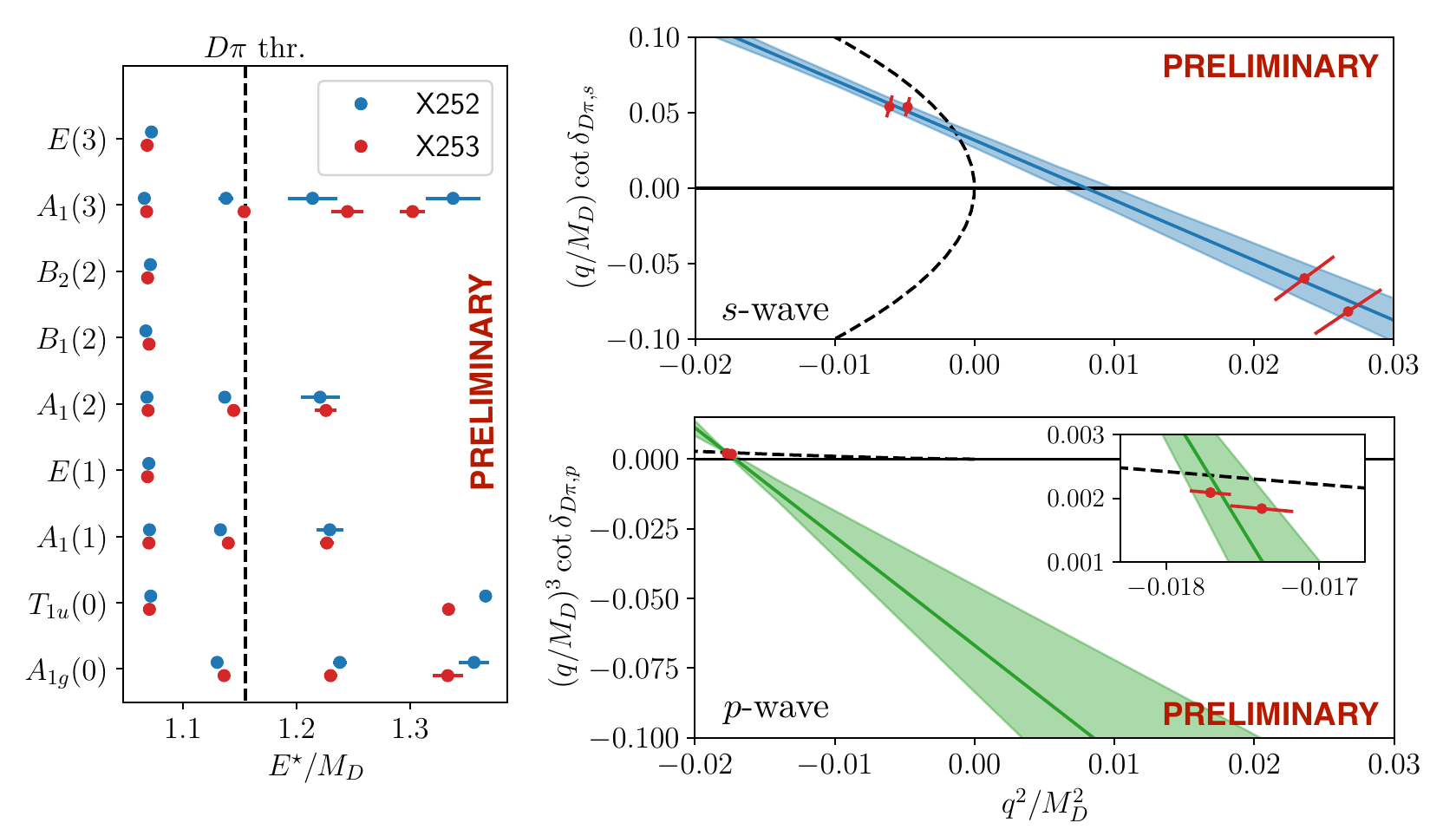}
\caption{The $D\pi$ spectrum on each ensemble is shown on the left. On the right, the $s$- and $p$-wave phase shift plots show the best fits described in the main text. The rest frame $A_{1g}$ and $T_{1u}$ levels included in the fit are displayed in red. The $D^*_0$ appears as a virtual bound state in $s$-wave, while the $D^*$ is a bound state in the $p$-wave, as expected. The inset zooms in on the intersection of the bound state condition (dashed line).}
\label{fig:Dpi_system}
\end{figure}

To constrain the $s$- and $p$-wave amplitudes for the $I=1/2$ $D\pi$ subchannel, a total of 38 levels were extracted, 19 each on the X252 and X253 ensembles. These were obtained on the $s$- and $p$-wave dominated $A_{1g}(0)$, $T_{1u}(0)$, $A_1(1)$, $A_1(2)$ and $A_1(3)$ irreps, with the ground states in the $p$-wave dominated $E(1)$, $B_1(2)$, $B_2(2)$, $E(3)$ irreps also included to provide extra constraints on the $D^*$ state. At this pion mass ($\sim 280$ MeV), the $D^*$ is expected to be bound, as confirmed by the spectrum shown on the left side of Figure \ref{fig:Dpi_system}.

A variety of phase shift parameterizations were used to perform the analysis. The forms chosen were modified ERE2s:
\begin{align}
q \cot \delta_{D\pi, s} &= \frac12 r_{0,s} \left(q^2 - q_{0,s}^2 \right) \,, \\
q^3 \cot \delta_{D\pi, p} &= \frac12 r_{0,p} \left(q^2 - q_{0,p}^2 \right) \,.
\end{align}
For $s$-wave, the effective range $r_{0,s}$ and zero crossing $q_{0,s}^2$ were used as fit parameters. For $p$-wave, the heavy-quark effective theory (HQET) $D^*D\pi$ coupling $g$ and the $D^*$ mass $M_{D^*}$ were used instead. These fix $r_{0,p}$ and $q_{0,p}^2$ uniquely by requiring that the $D^*$ pole position and residue match between the two expressions for the $p$-wave amplitude, as worked out in detail in Appendix D of ref.~\cite{Dawid:2024dgy}.

While the $D^*$ pole position, given by its mass $M_{D^*}$, seems very precisely constrained by the $D\pi$ data alone, its residue, set by the coupling $g$, is not. Therefore, we introduced a Bayesian prior $g = 0.55(10)$ to stabilize these preliminary fits. This value was chosen in line with ref.~\cite{Dawid:2024dgy}, close to the experimental value $\sim 0.57$ \cite{BaBar:2013zgp} and within the range of lattice-determined values $0.50-0.65$ \cite{Becirevic:2012pf, Can:2012tx}. We observed that the coupling and its error seem to effectively be set by the prior.

For the best fit of the 32 selected levels, the parameters were 
\begin{align} \nonumber
q_{0,s}^2/M_D^2 &= 0.0080(18)\,, \qquad r_{0,s} M_D = -7.91(80) \,, \\
M_{D^*}/M_D &= 1.07041(63) \,, \qquad gM_D = 0.542(96)
\end{align}
leading to $\chi^2 / {\rm d.o.f.} = 36.5/29 = 1.26 $. The best fits are shown on the right-hand side of Figure \ref{fig:Dpi_system}.

\subsection{Isospin-0 \ensuremath{DD^* + DD\pi} system}
\label{sec:DDs_fitting}

We turn now to the main $DD^* + DD\pi$ system. Here, 12 levels obtained in X252 were selected for the fitting, on the $T_{1g}(0)$, $A_{1u}(0)$, $A_2(1)$ and $A_2(2)$ irreps.

\begin{figure}[h]
\centering
\includegraphics[width=0.82\linewidth]{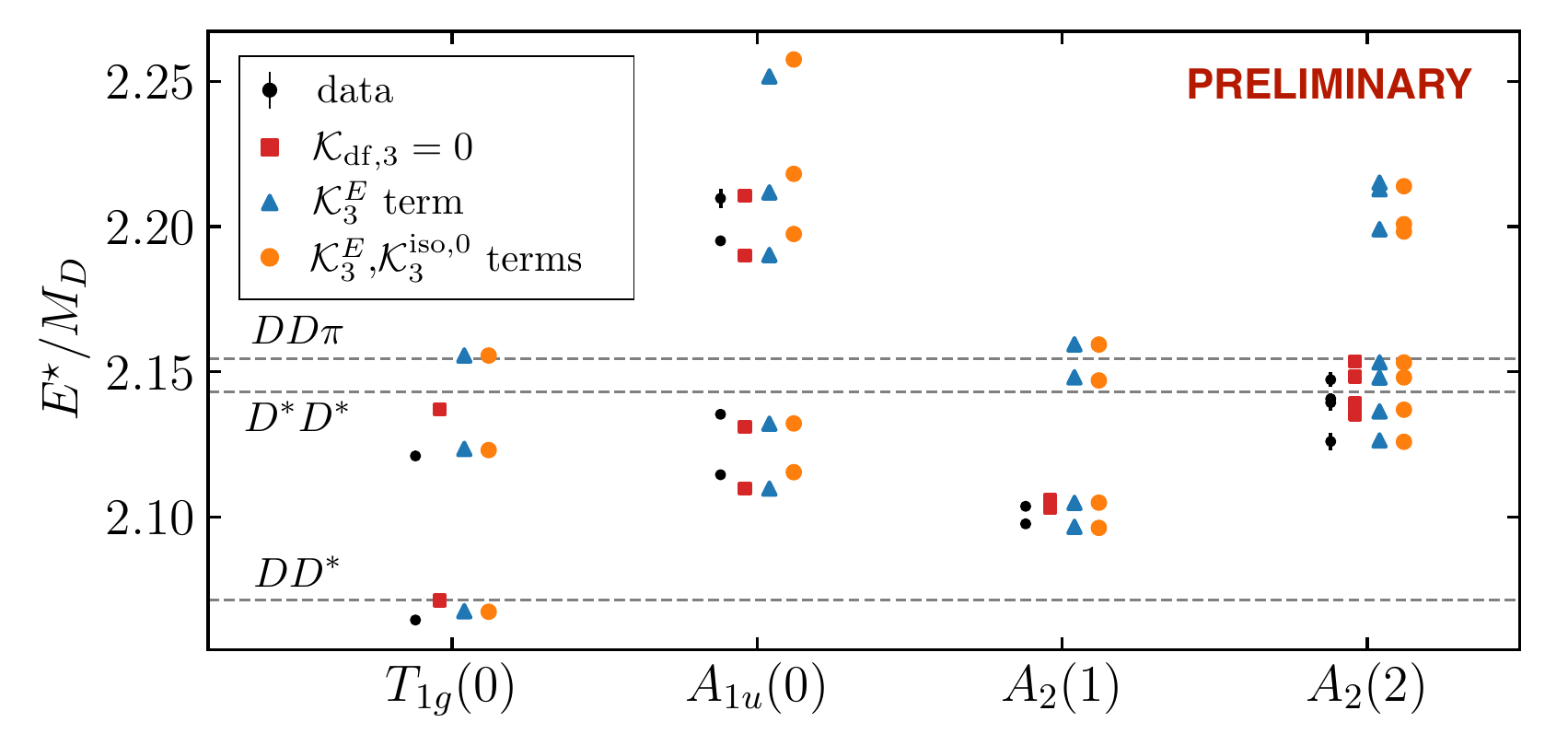}
\caption{Comparison of the selected $DD^* + DD\pi$ levels extracted on X252 (black) with the QC3 predictions for vanishing $\mathcal K_{\rm df, 3}$ (red), and the two fits described in the main text (blue and orange). The relevant two- and three-body thresholds are shown as dashed lines.}
\label{fig:3B_fits}
\end{figure}

A first check of the spectrum is provided by solving the QC3 of eq.~\eqref{eq:QC3_def}, with the three-body K-matrix $\widehat{\mathcal K}_{3,
\rm df}$ set to zero and the $DD$ and $D\pi$ two-body K-matrices entering $\widehat{F}_3$ fixed to the models and parameters obtained in sections \ref{sec:DD_fitting} and \ref {sec:Dpi_fitting}. The plot in Figure \ref{fig:3B_fits} shows the comparison of the data with the QC3 prediction -- although there is some correspondence between the data points and QC3 solutions, these give a rather poor description of the data with $\chi^2 \approx 247$. This is unsurprising given the observation in previous work \cite{Dawid:2024dgy} that a nonzero $\widehat{\mathcal K}_{3,\rm df}$ was needed to adequately describe the lattice data of ref.~\cite{Padmanath:2022cvl}.

In ref.~\cite{Dawid:2024dgy}, it was shown that the finite-volume spectrum predicted by the QC3 is most sensitive to the $\mathcal K_3^E$ term in the parameterization of eq.~\eqref{eq:K3_thr}. Therefore, an initial fit to the twelve levels with $\mathcal K_3^E$ as the single parameter was performed. This already shows considerable improvement, lowering the $\chi^2$ down to $\sim 104$, with 11 degrees of freedom. Going further, we can include a constant $\mathcal K_3^{{\rm iso}, 0}$ term, mostly impacting the $A_{1u}$ levels and leading to a further decrease to $\chi^2 \approx 66$, with 10 degrees of freedom.

Although more terms can be added to the $\mathcal K_{\rm df, 3}$ parameterization, we consider that the most promising way forward is to move towards simultaneous fits to the $DD$, $D\pi$ and $DD^*+DD\pi$ data, as has been done for three-body systems of pions and kaons in ref.~\cite{Draper:2023boj}. This way, the two-body subchannels can more directly inform the fit of the $DD\pi + DD^*$ data and vice-versa. This could prove helpful in resolving issues such as constraining the $D^*D\pi$ coupling $g$.

\section{Summary and outlook}
\label{sec:outlook}

We have presented preliminary results from a first lattice QCD study of the $T_{cc}^+$ tetraquark via a three-body approach \cite{Hansen:2024ffk}. This approach circumvents the left-hand cut issues that have affected previous investigations of the $T_{cc}^+$ via $DD^*$ scattering using the L\"uscher formalism, also treating three-body $DD\pi$ effects, which are crucial for future studies at physical or near-physical pion masses.

Finite-volume spectra for the $I=0$ $DD^* + DD\pi$ system and the two-particle subsystems $I=1$ $DD$ and $I=1/2$ $D\pi$ were extracted on the X252 CLS ensemble at $M_\pi \approx 280$ MeV, with $D\pi$ using also the X253 ensemble. The subchannel systems were analyzed individually to determine suitable models for the two-body K-matrices, required as inputs for the three-body quantization condition. The $DD^* + DD\pi$ spectrum was then fit using fixed two-body parameters and simple three-particle K-matrix parameterizations, leading to a reasonable but improvable description of the data.

In the near future, energies will be extracted on the X253 ensemble for all three systems, allowing us to probe more kinematics. The clear next step will then be to perform simultaneous fits to all three spectra, concurrently constraining the two- and three-body physics. With a reasonable determination of the three-body K-matrix, we can move on to solving the integral equations to obtain the $DD^*$ scattering amplitude and search for poles associated with the $T_{cc}^+$. The machinery for this step is largely in place from the previous work of ref.~\cite{Dawid:2024dgy}.

\section*{Acknowledgments}

The work of ABR and JB is supported by the European Research Council (ERC) consolidator grant StrangeScatt-101088506.
The work of FRL and MS was supported in part by the Swiss National Science Foundation (SNSF) through grant No.~200021-236432, and the Platform for Advanced Scientific Computing (PASC) project “ALPENGLUE”.
The work of SRS is supported in part by the U.S.~Department of Energy grant No.~DE-SC0011637. This work contributes to the goals of the USDOE ExoHad Topical Collaboration, contract DE-SC0023598. SMD acknowledges support from the contract. The research of AS is funded by the Deutsche Forschungsgemeinschaft (DFG, German Research Foundation) - Projektnummer 417533893/GRK2575 ``Rethinking Quantum Field Theory''.
C.J.M.~acknowledges support from the
U.S.~National Science Foundation (NSF) under award PHY-2514831. 

ABR thanks the Albert Einstein Center at the University of Bern for its
hospitality during a visit that significantly advanced this
work. 

Part of the calculations were performed on UBELIX (\url{https://www.id.unibe.ch/hpc}), the HPC cluster at the University of Bern. Part of the computations used a grant from the Swiss National Supercomputing Centre (CSCS) under project ID lp53 on Alps.
A small part of the calculations was also performed using resources provided by the Gauss Centre for Supercomputing e.V.\ (\url{www.gauss-centre.eu}) on JUWELS~\cite{juwels} at the Jülich Supercomputing Centre and on the HPC cluster Elysium of the Ruhr-Universit\"at Bochum, subsidized by the DFG (INST 213/1055-1).

\bibliographystyle{JHEP}
\bibliography{refs.bib}

@article{LHCb:2021vvq,
    author = "Aaij, Roel and others",
    collaboration = "LHCb",
    title = "{Observation of an exotic narrow doubly charmed tetraquark}",
    eprint = "2109.01038",
    archivePrefix = "arXiv",
    primaryClass = "hep-ex",
    reportNumber = "CERN-EP-2021-165, LHCb-PAPER-2021-031",
    doi = "10.1038/s41567-022-01614-y",
    journal = "Nature Phys.",
    volume = "18",
    number = "7",
    pages = "751--754",
    year = "2022"
}

@article{Lyu:2023xro,
    author = "Lyu, Yan and Aoki, Sinya and Doi, Takumi and Hatsuda, Tetsuo and Ikeda, Yoichi and Meng, Jie",
    title = "{Doubly Charmed Tetraquark $T_{cc}^+$ from Lattice QCD near Physical Point}",
    eprint = "2302.04505",
    archivePrefix = "arXiv",
    primaryClass = "hep-lat",
    reportNumber = "RIKEN-iTHEMS-Report-23, YITP-23-14",
    doi = "10.1103/PhysRevLett.131.161901",
    journal = "Phys. Rev. Lett.",
    volume = "131",
    number = "16",
    pages = "161901",
    year = "2023"
}

@article{LHCb:2021auc,
    author = "Aaij, Roel and others",
    collaboration = "LHCb",
    title = "{Study of the doubly charmed tetraquark $T_{cc}^{+}$}",
    eprint = "2109.01056",
    archivePrefix = "arXiv",
    primaryClass = "hep-ex",
    reportNumber = "CERN-EP-2021-169, LHCb-PAPER-2021-032",
    doi = "10.1038/s41467-022-30206-w",
    journal = "Nature Commun.",
    volume = "13",
    number = "1",
    pages = "3351",
    year = "2022"
}

@article{Padmanath:2022cvl,
    author = "Padmanath, M. and Prelovsek, S.",
    title = "{Signature of a Doubly Charm Tetraquark Pole in DD* Scattering on the Lattice}",
    eprint = "2202.10110",
    archivePrefix = "arXiv",
    primaryClass = "hep-lat",
    reportNumber = "MITP/22-018",
    doi = "10.1103/PhysRevLett.129.032002",
    journal = "Phys. Rev. Lett.",
    volume = "129",
    number = "3",
    pages = "032002",
    year = "2022"
}

@article{Chen:2022vpo,
    author = "Chen, Siyang and Shi, Chunjiang and Chen, Ying and Gong, Ming and Liu, Zhaofeng and Sun, Wei and Zhang, Renqiang",
    title = "{$T_{cc}^+(3875)$ relevant $DD^*$ scattering from $N_f=2$ lattice QCD}",
    eprint = "2206.06185",
    archivePrefix = "arXiv",
    primaryClass = "hep-lat",
    doi = "10.1016/j.physletb.2022.137391",
    journal = "Phys. Lett. B",
    volume = "833",
    pages = "137391",
    year = "2022"
}

@article{Collins:2024sfi,
    author = "Collins, Sara and Nefediev, Alexey and Padmanath, M. and Prelovsek, Sasa",
    title = "{Toward the quark mass dependence of ${T_{cc}^+}$ from lattice QCD}",
    eprint = "2402.14715",
    archivePrefix = "arXiv",
    primaryClass = "hep-lat",
    reportNumber = "IMSc/24/01",
    doi = "10.1103/PhysRevD.109.094509",
    journal = "Phys. Rev. D",
    volume = "109",
    number = "9",
    pages = "094509",
    year = "2024"
}

@article{Dawid:2024oey,
    author = "Dawid, Sebastian M. and Jackura, Andrew W. and Szczepaniak, Adam P.",
    title = "{Finite-volume quantization condition from the N/D representation}",
    eprint = "2411.15730",
    archivePrefix = "arXiv",
    primaryClass = "hep-lat",
    reportNumber = "JLAB-THY-24-4216",
    doi = "10.1016/j.physletb.2025.139442",
    journal = "Phys. Lett. B",
    volume = "864",
    pages = "139442",
    year = "2025"
}

@article{Whyte:2024ihh,
    author = "Whyte, Travis and Wilson, David J. and Thomas, Christopher E.",
    collaboration = "Hadron Spectrum",
    title = "{Near-threshold states in coupled DD*-D*D* scattering from lattice QCD}",
    eprint = "2405.15741",
    archivePrefix = "arXiv",
    primaryClass = "hep-lat",
    doi = "10.1103/PhysRevD.111.034511",
    journal = "Phys. Rev. D",
    volume = "111",
    number = "3",
    pages = "034511",
    year = "2025"
}

@article{Prelovsek:2025vbr,
    author = "Prelovsek, Sasa and Ortiz-Pacheco, Emmanuel and Collins, Sara and Leskovec, Luka and Padmanath, M. and Vujmilovic, Ivan",
    title = "{Doubly heavy tetraquarks from lattice QCD: Incorporating diquark-antidiquark operators and the left-hand cut}",
    eprint = "2504.03473",
    archivePrefix = "arXiv",
    primaryClass = "hep-lat",
    doi = "10.1103/rlgp-c9tb",
    journal = "Phys. Rev. D",
    volume = "112",
    number = "1",
    pages = "014507",
    year = "2025"
}

@article{Meng:2023bmz,
    author = "Meng, Lu and Baru, Vadim and Epelbaum, Evgeny and Filin, Arseniy A. and Gasparyan, Ashot M.",
    title = "{Solving the left-hand cut problem in lattice QCD: $T_{cc}(3875)^+$ from finite volume energy levels}",
    eprint = "2312.01930",
    archivePrefix = "arXiv",
    primaryClass = "hep-lat",
    doi = "10.1103/PhysRevD.109.L071506",
    journal = "Phys. Rev. D",
    volume = "109",
    number = "7",
    pages = "L071506",
    year = "2024"
}

@article{Du:2023hlu,
    author = "Du, Meng-Lin and Filin, Arseniy and Baru, Vadim and Dong, Xiang-Kun and Epelbaum, Evgeny and Guo, Feng-Kun and Hanhart, Christoph and Nefediev, Alexey and Nieves, Juan and Wang, Qian",
    title = "{Role of Left-Hand Cut Contributions on Pole Extractions from Lattice Data: Case Study for Tcc(3875)+}",
    eprint = "2303.09441",
    archivePrefix = "arXiv",
    primaryClass = "hep-ph",
    doi = "10.1103/PhysRevLett.131.131903",
    journal = "Phys. Rev. Lett.",
    volume = "131",
    number = "13",
    pages = "131903",
    year = "2023"
}

@article{Raposo:2023oru,
    author = "Raposo, Andr{\'e} Bai{\~a}o and Hansen, Maxwell T.",
    title = "{Finite-volume scattering on the left-hand cut}",
    eprint = "2311.18793",
    archivePrefix = "arXiv",
    primaryClass = "hep-lat",
    doi = "10.1007/JHEP08(2024)075",
    journal = "JHEP",
    volume = "08",
    pages = "075",
    year = "2024"
}

@article{Hansen:2024ffk,
    author = "Hansen, Maxwell T. and Romero-L{\'o}pez, Fernando and Sharpe, Stephen R.",
    title = "{Incorporating DD{\ensuremath{\pi}} effects and left-hand cuts in lattice QCD studies of the T$_{cc}$(3875)$^{+}$}",
    eprint = "2401.06609",
    archivePrefix = "arXiv",
    primaryClass = "hep-lat",
    reportNumber = "MIT-CTP/5667",
    doi = "10.1007/JHEP06(2024)051",
    journal = "JHEP",
    volume = "06",
    pages = "051",
    year = "2024"
}

@article{Bubna:2024izx,
    author = {Bubna, Rishabh and Hammer, Hans-Werner and M{\"u}ller, Fabian and Pang, Jin-Yi and Rusetsky, Akaki and Wu, Jia-Jun},
    title = {{L{\"u}scher equation with long-range forces}},
    eprint = "2402.12985",
    archivePrefix = "arXiv",
    primaryClass = "hep-lat",
    doi = "10.1007/JHEP05(2024)168",
    journal = "JHEP",
    volume = "05",
    pages = "168",
    year = "2024"
}

@article{Raposo:2025dkb,
    author = "Raposo, Andr{\'e} Bai{\~a}o and Brice{\~n}o, Ra{\'u}l A. and Hansen, Maxwell T. and Jackura, Andrew W.",
    title = "{Extracting scattering amplitudes for arbitrary two-particle systems with one-particle left-hand cuts via lattice QCD}",
    eprint = "2502.19375",
    archivePrefix = "arXiv",
    primaryClass = "hep-lat",
    doi = "10.1007/JHEP06(2025)186",
    journal = "JHEP",
    volume = "06",
    pages = "186",
    year = "2025"
}

@article{Dawid:2024dgy,
    author = "Dawid, Sebastian M. and Romero-L{\'o}pez, Fernando and Sharpe, Stephen R.",
    title = "{Finite- and infinite-volume study of DD{\ensuremath{\pi}} scattering}",
    eprint = "2409.17059",
    archivePrefix = "arXiv",
    primaryClass = "hep-lat",
    reportNumber = "MIT-CTP/5774",
    doi = "10.1007/JHEP01(2025)060",
    journal = "JHEP",
    volume = "01",
    pages = "060",
    year = "2025"
}

@article{Draper:2023boj,
    author = "Draper, Zachary T. and Hanlon, Andrew D. and H{\"o}rz, Ben and Morningstar, Colin and Romero-L{\'o}pez, Fernando and Sharpe, Stephen R.",
    title = "{Interactions of {\ensuremath{\pi}}K, {\ensuremath{\pi}}{\ensuremath{\pi}}K and KK{\ensuremath{\pi}} systems at maximal isospin from lattice QCD}",
    eprint = "2302.13587",
    archivePrefix = "arXiv",
    primaryClass = "hep-lat",
    reportNumber = "MIT-CTP/5536",
    doi = "10.1007/JHEP05(2023)137",
    journal = "JHEP",
    volume = "05",
    pages = "137",
    year = "2023"
}

@article{Blanton:2021mih,
    author = "Blanton, Tyler D. and Sharpe, Stephen R.",
    title = "{Three-particle finite-volume formalism for {\ensuremath{\pi^+\pi^+K^+}} and related systems}",
    eprint = "2105.12094",
    archivePrefix = "arXiv",
    primaryClass = "hep-lat",
    doi = "10.1103/PhysRevD.104.034509",
    journal = "Phys. Rev. D",
    volume = "104",
    number = "3",
    pages = "034509",
    year = "2021"
}

@article{Stump:2025owq,
    author = "Stump, Andres and Green, Jeremy R.",
    title = "{Position-space sampling for local multiquark operators in lattice QCD using distillation and the importance of tetraquark operators for $T_{cc}(3875)^+$}",
    eprint = "2510.26459",
    archivePrefix = "arXiv",
    primaryClass = "hep-lat",
    reportNumber = "HU-EP-25/35-RTG, DESY-25-140",
    month = "10",
    year = "2026",
    journal = "Phys. Rev. D",
    pages = "(to be published)"
}

@software{quda_laph,
  author       = "{Cosmon Collaboration}",
  title        = "{quda laph: Lattice QCD computations with stochastic Laph and distillation using the QUDA library
  }",
  year         = {2025},
  url          = "https://github.com/cosmon-collaboration/quda_laph",
  version      = {v0.9}
}

@article{BaBar:2013zgp,
    author = "Lees, J. P. and others",
    collaboration = "BaBar",
    title = "{Measurement of the $D^*(2010)^+$ natural line width and the $D^*(2010)^+ - D^0$ mass difference}",
    eprint = "1304.5009",
    archivePrefix = "arXiv",
    primaryClass = "hep-ex",
    reportNumber = "BABAR-PUB-12-032, SLAC-PUB-15374",
    doi = "10.1103/PhysRevD.88.052003",
    journal = "Phys. Rev. D",
    volume = "88",
    number = "5",
    pages = "052003",
    year = "2013",
    note = "[Erratum: Phys.Rev.D 88, 079902 (2013)]"
}

@article{Becirevic:2012pf,
    author = "Becirevic, Damir and Sanfilippo, Francesco",
    title = "{Theoretical estimate of the $D^* \to D\pi$ decay rate}",
    eprint = "1210.5410",
    archivePrefix = "arXiv",
    primaryClass = "hep-lat",
    reportNumber = "LPT-ORSAY-12-105",
    doi = "10.1016/j.physletb.2013.03.004",
    journal = "Phys. Lett. B",
    volume = "721",
    pages = "94--100",
    year = "2013"
}

@article{Can:2012tx,
    author = "Can, K. U. and Erkol, G. and Oka, M. and Ozpineci, A. and Takahashi, T. T.",
    title = "{Vector and axial-vector couplings of D and D* mesons in 2+1 flavor Lattice QCD}",
    eprint = "1210.0869",
    archivePrefix = "arXiv",
    primaryClass = "hep-lat",
    doi = "10.1016/j.physletb.2012.12.050",
    journal = "Phys. Lett. B",
    volume = "719",
    pages = "103--109",
    year = "2013"
}

@article{HadronSpectrum:2009krc,
    author = "Peardon, Michael and Bulava, John and Foley, Justin and Morningstar, Colin and Dudek, Jozef and Edwards, Robert G. and Joo, Balint and Lin, Huey-Wen and Richards, David G. and Juge, Keisuke Jimmy",
    collaboration = "Hadron Spectrum",
    title = "{A Novel quark-field creation operator construction for hadronic physics in lattice QCD}",
    eprint = "0905.2160",
    archivePrefix = "arXiv",
    primaryClass = "hep-lat",
    reportNumber = "JLAB-THY-09-985",
    doi = "10.1103/PhysRevD.80.054506",
    journal = "Phys. Rev. D",
    volume = "80",
    pages = "054506",
    year = "2009"
}

@article{Hanhart:2025bun,
    author = "Hanhart, C.",
    title = "{Hadronic molecules and multiquark states}",
    eprint = "2504.06043",
    archivePrefix = "arXiv",
    primaryClass = "hep-ph",
    month = "4",
    year = "2025"
}

@article{juwels,
    author = {{J{\"u}lich Supercomputing Centre}},
    title = {{JUWELS Cluster and Booster}: Exascale Pathfinder with Modular Supercomputing Architecture at {Jülich Supercomputing Centre}},
    journal = {J. Large-Scale Res. Facil.},
    pages  = {A183},
    volume = {7},
    doi = {10.17815/jlsrf-7-183},
    year = {2021}
}

\end{document}